# An Analytical Study on Fractional Fokker-Planck Equation by Homotopy Analysis Transform Method

Jitendra Singh[*]
Department of Mathematics, Central University of Bihar, Patna-800014, Bihar.

[*] Corresponding author. Tel.: +91-06122784172.
E-mail address: jitendrasingh@cub.ac.in (Jitendra Singh)

**Abstract:** In this article, we introduce an analytical method, namely Homotopy Analysis Transform Method (HATM) which is a combination of Homotopy Analysis Method (HAM) and Laplace Decomposition Method (LDM). This scheme is simple to apply linear and nonlinear fractional differential equations and having less computational work in comparison of other exiting methods. The most useful advantage of this method is to solve the fractional nonlinear differential equation without using Adomian polynomials and He's polynomials for the computation of nonlinear terms. The proposed method has no linearization and restrictive assumptions throughout the process. Presently, homotopy analysis transform method is used to solve time fractional Fokker-Planck equation and similar equations. The series solution obtained by HATM converges very fast. A good agreement between the obtained solution and some well-known results has been demonstrated.

## 1. Introduction:

Fractional differential equations have long history. These equations have demonstrated a considerable interest both in mathematics and in applications in recent years. They have been used in modelling of many physical and chemical processes in engineering [1-3]. The aim of using fractional calculus facilitates the description of some problems which are not easy described by ordinary calculus due to modelling error. The advantage of the fractional order system is that they allow greater degree of freedom in the model. To find the solution of fractional differential equation is not easy due to many fundamental law of calculus having integer order derivatives is not hold good in case of fractional derivatives such as product rule etc. Saadatmandi and Dehghan [4] have solved numerically the fractional order differential equation using Legendre operational matrix. Hesameddini et. al. [5] solved nonlinear fractional differential equation with a coupling method of Homotopy technique and Laplace transform. Partial differential equations of fractional order are often very complicated to be exactly solved and even if an exact solution is obtainable, the required calculations may be too complicated to be practical. Finding accurate and efficient methods for solving fractional differential equations becomes an important task. We used an analytical method homotopy analysis transform method (HATM) for the solution of time Fractional Fokker-Planck equation in closed form.

The Fokker-Plank equation manly generates in natural science likes quantum optics, chemical physics, fluid dynamic, theoretical biology, solid-state physics and circuit theory. Fokker and Planck [6] used this equation to describe the Brownian motion of particles. The fractional order system of the Fokker-Plank equation are obtained by replacing the first order time derivative term by fractional order $\alpha \, (0 < \alpha \leq 1)$. A generalized time fractional Fokker-Plank equation in general form to n variables $x_1, x_2, \ldots\ldots x_n$ may be stated as,

$$\frac{\partial^\alpha u}{\partial t^\alpha} = \left[ -\sum_{i=1}^{n} \frac{\partial}{\partial x_i} A_i(x,t) + \sum_{i=1}^{n}\sum_{j=1}^{n} \frac{\partial^2}{\partial x_i \partial x_j} B_{i,j}(x,t) \right] u(x,t) \qquad (0 < \alpha \leq 1) \qquad (1)$$

With the initial condition:
$$u(x,0) = f(x), \qquad x = (x_1, x_2, \ldots x_n) \in \mathbb{R}^n. \tag{2}$$

Eq. (1-2) is an equation of motion for the distribution function $u(x,t)$, which is called the time fractional forward Kolmogorov equation, where $B_{i,j}(x,t) > 0$ is called the diffusion coefficient and $A_i(x,t)$ is the drift coefficient. The similar partial differential equation, which might know as the time fractional backward kolmogorov equation, can be written as [6],

$$\frac{\partial^\alpha u}{\partial t^\alpha} = \left[ -A_i(x,t) \sum_{i=1}^n \frac{\partial}{\partial x_i} + B_{i,j}(x,t) \sum_{i=1}^n \sum_{j=1}^n \frac{\partial^2}{\partial x_i \partial x_j} \right] u(x,t) \qquad (0 < \alpha \leq 1), \tag{3}$$

The nonlinear Fokker-Planck equation provides useful mathematical models in different areas such as plasma physics, surface physics, population dynamics, biophysics, engineering, nonlinear hydrodynamics, polymer physics, psychology and marketing [7]. A generalized nonlinear time fractional Fokker-Planck equation in n variables, $x_1, x_2, \ldots x_n$ may be written as

$$\frac{\partial^\alpha u}{\partial t^\alpha} = \left[ -\sum_{i=1}^n \frac{\partial}{\partial x_i} A_i(x,t,u) + \sum_{i=1}^n \sum_{j=1}^n \frac{\partial^2}{\partial x_i \partial x_j} B_{i,j}(x,t,u) \right] u(x,t) \qquad (0 < \alpha \leq 1), \tag{4}$$

The numerical and approximate analytical solutions of the Fokker-Planck equation have been studied since the work of Risken [6]. There are several methods in literature which are used to solve this equation [8-16]. Some methods are based on transformation technique and some other give the solution in series form which converges to the exact solution. Variational iteration and Homotopy perturbation method [17] and the Adomian decomposition method [18] have been used to get exact and approximate analytical solutions. Recently S. Hesam et. al. [19] successfully solved the Fokker-Planck equation by differential transform method. Motivated by the above work, In this article, authors are concerned to solve time fractional Fokker-Planck equations and time fractional kolmogorov equation. We are effectively found the solution for linear and nonlinear time fractional Fokker-Planck equations by using homotopy analysis transform method. Homotopy analysis transform method is combination of Homotopy analysis method [20-24] with Laplace decomposition method [25-28]. In the HAM, how to use the initial approximation, auxiliary linear operator, auxiliary function and the convergence controlling auxiliary parameter are given in [29-31]. The advantage of the proposed method is its capability of combining two powerful methods for obtaining rapid convergent series for the solution of fractional nonlinear partial differential equations. Khan et. al. [32] used this method to solve the famous Blasius flow equation on a semi infinite domain. This paper consider the effectiveness of the homotopy analysis transform method in solving nonlinear time fractional Fokker-Planck equation with different fractional Brownian motion and also for the standard motion. To the best of author's knowledge these problems with fractional time derivative have not yet been solved by any researcher.

The Paper is organized as follows, In section 2, the basic idea of the HATM, section 3, contains basic definition of Caputo order fractional derivative, section 4, application of HATM on some examples with analytical solutions in one and two dimensional cases, in last, section 5 have conclusions.

**2. Homotopy Analysis Transform Method(HATM)**

Consider equation $N(u(x,t)) = g(x,t)$, where $N$ represents a general nonlinear ordinary or partial differential operator including both linear and nonlinear terms. The linear terms is decomposed into $L + R$, where $L$ is the highest order linear operator and $R$ is the remaining of the linear operator. Thus, the equation can be written as

$$Lu + Ru + Nu = g(x,t), \tag{5}$$

Where $Nu$ indicates the nonlinear terms. By applying Laplace transform on both sides of Eq. (5), we get

$$\pounds[Lu + Ru + Nu = g(x,t)]. \tag{6}$$

Using the differential property of Laplace transform, we have

$$s^n \pounds[u] - \sum_{k=1}^{n} s^{k-1} u^{(n-k)}(0) + \pounds[Ru] + \pounds[Nu] = \pounds[g(x,t)], \tag{7}$$

On simplifying

$$\pounds[u] - \frac{1}{s^n} \sum_{k=1}^{n} s^{k-1} u^{(n-k)}(0) + \frac{1}{s^n}(\pounds[Ru] + \pounds[Nu]) - \frac{1}{s^n}(\pounds[g(x,t)]) = 0, \tag{8}$$

We define the nonlinear operator

$$N[\phi(x,t;q)] = \pounds[\phi(x,t;q)] - \frac{1}{s^n} \sum_{k=1}^{n} s^{k-1} \phi^{(n-k)}(x,0;q) + \frac{1}{s^n}(\pounds[R(\phi(x,t;q))] + \pounds[N(\phi(x,t;q))])$$

$$- \frac{1}{s^n}(\pounds[g(x,t)]). \tag{9}$$

Where $q \in [0,1]$ and $\phi(x,t;q)$ is real function of $x, t$ and $q$. We construct a homotopy as follows,

$$(1-q)\pounds[\phi(x,t;q) - u_0(x,t)] = \hbar H(x,t) q N[\phi(x,t;q)], \tag{10}$$

Where $\pounds$ denotes the Laplace transform, $q \in [0,1]$ is the embedding parameter, $\hbar$ is a non-zero auxiliary parameter, $H(x,t) \neq 0$ is an auxiliary function, $u_0(x,t)$ is an initial guess of $u(x,t)$ and $\phi(x,t;q)$ is an unknown function. Obviously, when $q = 0$ and $q = 1$, it holds,

$$\phi(x,t;0) = u_0(x,t), \qquad \phi(x,t;1) = u(x,t). \tag{11}$$

respectively. Thus, as $q$ increases from 0 to 1, the solution $\phi(x,t;q)$ varies from the initial guess $u_0(x,t)$ to the solution $u(x,t)$. Expanding $\phi(x,t;q)$ in Taylor series with respect to $q$, we have

$$\phi(x,t;q) = u_0(x,t) + \sum_{m=1}^{\infty} u_m(x,t) q^m, \tag{12}$$

Where

$$u_m(x,t) = \frac{1}{m!} \frac{\partial^m \phi(x,t;q)}{\partial q^m}\bigg|_{q=0}. \tag{13}$$

If the auxiliary linear operator (L), the initial guess $u_0(x,t)$, the auxiliary parameter $\hbar$ and the auxillary function $H(x,t)$ are properly chosen, series (12) converges at $q = 1$, then we have

$$u(x,t) = u_0(x,t) + \sum_{m=1}^{\infty} u_m(x,t), \tag{14}$$

Which must be one of solutions of original nonlinear equations. According to definition (14), the governing equation can be deduced from the zero-order deformation (10). Define the vector

$$\vec{u}(x,t) = \{u_0(x,t), u_1(x,t), u_2(x,t) \ldots \ldots u_m(x,t)\}. \tag{15}$$

Differentiating Eq. (10) $m$ times with respect to the embedding parameter $q$ and then setting $q = 0$ and finally dividing them by $m!$, we have the so called mth-order deformation equation

$$\pounds[u_m(x,t) - \chi_m u_{m-1}(x,t)] = \hbar H(x,t) R_m(\vec{u}_{m-1}). \tag{16}$$

Applying inverse Laplace transform, we have

$$u_m(x,t) = \chi_m u_{m-1}(x,t) + \hbar \pounds^{-1}[H(x,t) R_m(\vec{u}_{m-1})], \tag{17}$$

Where

$$R_m(\vec{u}_{m-1}) = \frac{1}{(m-1)!} \frac{\partial^{m-1} N(x,t;q)}{\partial q^{m-1}} \bigg|_{q=0}, \tag{18}$$

And

$$\chi_m = \begin{cases} 0, & m \leq 1 \\ 1, & m > 1 \end{cases}. \tag{19}$$

### 3. Basic definition of fractional derivative in Caputo sense

There are various definitions of fractional integration and differentiation, such as Grunwald-Letnikov's definition and Riemann-Liouville's definition. The Riemann-Liouville derivative has certain disadvantages when trying to model real-world phenomena with fractional differential equations. Therefore, we shall introduce a modified fractional differential operator $D^\alpha$ proposed by Caputo in his work on the theory of viscoelasticity [33-34].

**Definition 3.1** The Caputo definition of the fractional-order derivative is defined as

$$D^\alpha f(x) = \frac{1}{\Gamma(n-\alpha)} \int_0^x \frac{f^{(n)}(t)}{(x-t)^{\alpha+1-n}} dt, \quad n-1 < \alpha \leq n, \; n \in \mathbb{N}, \tag{20}$$

Where $\alpha > 0$ is the order of the derivative and $n$ is the smallest integer greater than $\alpha$. For the Caputo derivative, we have

$$D^\alpha C = 0, \quad (C \text{ is a constant}), \tag{21}$$

$$D^\alpha x^\beta = \begin{cases} 0, & \text{for } \beta \in \mathbb{N}_0 \text{ and } \beta < \lceil \alpha \rceil, \\ \frac{\Gamma(1+\beta)}{\Gamma(1+\beta-\alpha)} x^{\beta-\alpha}, & \text{for } \beta \in \mathbb{N}_0 \text{ and } \beta \geq \lceil \alpha \rceil \text{ or } \beta \notin \mathbb{N} \text{ and } \beta > \lfloor \alpha \rfloor. \end{cases} \tag{22}$$

Where $\lceil \alpha \rceil$ is ceiling function and $\lfloor \alpha \rfloor$ is floor function. $\mathbb{N} = \{1,2,.......\}$ and $\mathbb{N}_0 = \{0,1,2,....\}$ Recall that for $\alpha \in \mathbb{N}$, the Caputo differential operator coincides with the usual differential operator of an integer order. Similar to the integer-order differentiation, the Caputo fractional differentiation is a linear operation

$$D^\alpha(\lambda f(x) + \mu g(x)) = \lambda D^\alpha f(x) + \mu D^\mu g(x), \tag{23}$$

Where $\lambda$ and $\mu$ are constants. In this paper, the fractional derivatives are considered in the Caputo sense. The reason for adopting the Caputo definition, as pointed by [35-36].

**Definition 3.2** The Laplace transform $\pounds[f(t); s]$ of the Caputo fractional derivative is given by

$$\pounds[D_t^\alpha f(t); s] = s^\alpha \pounds[f(t)] - \sum_{k=0}^{m-1} s^{\alpha-k-1} f^{(k)}(0), \quad m-1 < \alpha \leq m. \tag{24}$$

### 4. Numerical experiments

In this section, we shall illustrate the homotopy analysis transform technique by some examples.

**Example 4.1** Consider Eq. (1) and Eq. (2) with $n=1$ and $x_1 = x$,

$$A_1(x,t) = -1, \; B_{1,1}(x,t)=1, \; f(x) = x, \quad x \in \mathbb{R}. \tag{25}$$

Then Eq. (1) and Eq. (2) converted in to the following form:

$$\frac{\partial^\alpha u}{\partial t^\alpha} = \frac{\partial u}{\partial x} + \frac{\partial^2 u}{\partial x^2} \tag{26}$$

With the initial condition:

$$u(x,0) = x, \quad x \in \mathbb{R}. \tag{27}$$

According to HATM, we take Laplace transform of Eq. (26) and using the initial condition Eq. (27) and Eq. (24) for the Laplace transform of fractional order derivative. We have

$$\pounds[u(x,t)] - \frac{1}{s}u(x,0) - \frac{1}{s^\alpha}\pounds[u_x(x,t)] - \frac{1}{s^\alpha}\pounds[u_{xx}(x,t)] = 0, \tag{28}$$

We define the nonlinear operator

$$N[\phi(x,t;q)] = \pounds[\phi(x,t;q)] - \frac{1}{s}u(x,0) - \frac{1}{s^\alpha}\pounds[\phi_x(x,t;q)] - \frac{1}{s^\alpha}\pounds[\phi_{xx}(x,t;q)] = 0, \tag{29}$$

The series solution of the problem is given by

$$u(x,t) = u_0(x,t) + \sum_{m=1}^{\infty} u_m(x,t), \tag{30}$$

Where

$$u_m(x,t) = \frac{1}{m!} \frac{\partial^m \phi(x,t;q)}{\partial q^m}\bigg|_{q=0}. \tag{31}$$

The mth-order deformation equation, We obtain as in Eq. (16) and applying inverse Laplace Transform and using auxiliary function $H(x,t) = 1$, we have

$$u_m(x,t) = \chi_m u_{m-1}(x,t) + \hbar \pounds^{-1}[R_m(\vec{u}_{m-1})]. \quad (m \geq 1), \tag{32}$$

Where,

$$R_m(\vec{u}_{m-1}) = \pounds[u_{m-1}] - \frac{(1-\chi_m)}{s} u(x,0) - \frac{1}{s^\alpha}\pounds[(u_{m-1})_x] - \frac{1}{s^\alpha}\pounds[(u_{m-1})_{xx}], \tag{33}$$

and

$$\chi_m = \begin{cases} 0, & m \leq 1 \\ 1, & m > 1 \end{cases}. \tag{34}$$

Using the Eq. (27-34), we have the following HATM solution

$$u_0(x,t) = u(x,0) = x, \tag{35}$$

$$u_1(x,t) = -\frac{\hbar t^\alpha}{\Gamma(\alpha+1)}, \tag{36}$$

$$u_2(x,t) = -\frac{\hbar(\hbar+1)}{\Gamma(\alpha+1)} t^\alpha, \tag{37}$$

$$u_3(x,t) = -\frac{\hbar(\hbar+1)^2}{\Gamma(\alpha+1)} t^\alpha, \tag{38}$$

............

Now, we obtain the series solution from Eq. (30), using the Eq. (35-38). For getting the solution in closed form taking $\hbar = -1$, and $\alpha = 1$, we have

$$u(x,t) = x + t. \tag{39}$$

Which is exact solution of the Eq. (26-27) when $\alpha = 1$. This results show that it is hold good in case of $\alpha = 1$. In case of fractional differential equation, We may also find the other solution of Eq. (26-27) with the help of Eq. (30) using the Eq. (35-38) for different values of $\alpha$.

**Example 4.2** Consider Eq. (1) and Eq. (2) with $n = 1$ and $x_1 = x$,

$A_1(x,t) = e^t \coth(x)\cosh(x) + e^t \sinh(x) - \coth(x)$, $B_{1,1}(x,t) = e^t \cosh(x)$,

$f(x) = \sinh(x)$, $x \in \mathbb{R}$.

Using HATM we obtain the following relations:

$$N[\phi(x,t;q)] = \pounds[\phi(x,t;q)] - \frac{1}{s}u(x,0) - \frac{\cosh(x)}{s^\alpha}\pounds[e^t\phi_{xx}(x,t;q)] + \frac{\cosech(x)}{s^\alpha}\pounds[e^t\phi_x(x,t;q)]$$
$$+ \frac{\cosh x}{s^\alpha}\pounds[e^t\phi(x,t;q)] - \frac{\coth(x)\cosech(x)}{s^\alpha}\pounds[e^t\phi(x,t;q)] \quad (40)$$
$$- \frac{\coth(x)}{s^\alpha}\pounds[\phi_x(x,t;q)] + \frac{\cosech^2(x)}{s^\alpha}\pounds[\phi(x,t;q)],$$

$$R_m[\vec{u}_{m-1}(x,t)] = \pounds[u_{m-1}] - \frac{(1-\chi_m)}{s}u(x,0) - \frac{\cosh(x)}{s^\alpha}\pounds[e^t(u_{m-1})_{xx}] + \frac{\cosech(x)}{s^\alpha}\pounds[e^t(u_{m-1})_x]$$
$$+ \frac{\cosh x}{s^\alpha}\pounds[e^t u_{m-1}] - \frac{\coth(x)\cosech(x)}{s^\alpha}\pounds[e^t u_{m-1}] - \frac{\coth(x)}{s^\alpha}\pounds[(u_{m-1})_x] \quad (41)$$
$$+ \frac{\cosech^2(x)}{s^\alpha}\pounds[u_{m-1}],$$

The mth-order deformation equation, using auxiliary function $H(x,t)=1$, we have
$$u_m(x,t) = \chi_m u_{m-1}(x,t) + \hbar\pounds^{-1}[R_m(\vec{u}_{m-1})]. \quad (m \geq 1), \quad (42)$$

Using the Eq. (41-42), we have the following approximations
$$u_0(x,t) = u(x,0) = \sinh(x), \quad (43)$$
$$u_1(x,t) = t\sinh(x), \quad (44)$$
$$u_2(x,t) = \sinh(x)\frac{t^2}{2!} \quad (45)$$
$$u_3(x,t) = \sinh(x)\frac{t^3}{3!} \quad (46)$$
................

Now, we obtain the series solution from Eq. (14), using the Eq. (43-46). For getting the solution in closed form taking $\hbar = -1,$ and $\alpha = 1,$ we have
$$u(x,t) = \sinh(x)e^t \quad (47)$$
Which is exact solution of the given problem, when $\alpha = 1$.

**Example 4.3** consider the backward Kolmogorov Eq. (3) with $n=1, x_1 = x$
$A_1(x,t) = -(x+1),$
$B_{1,1}(x,t) = e^t x^2,$
$f(x) = x+1, \quad x \in \mathbb{R}.$
applying the HATM, we have
$$R_m(\vec{u}_{m-1}) = \pounds[u_{m-1}] - \frac{(1-\chi_m)}{s}u(x,0) - \frac{(1+x)}{s^\alpha}\pounds[(u_{m-1})_x] - \frac{x^2}{s^\alpha}\pounds[e^t(u_{m-1})_{xx}], \quad (48)$$
The mth-order deformation equation using auxiliary function $H(x,t)=1$, we have
$$u_m(x,t) = \chi_m u_{m-1}(x,t) + \hbar\pounds^{-1}[R_m(\vec{u}_{m-1})]. \quad (m \geq 1), \quad (49)$$
Using the Eq. (48-49), we have the following approximations
$$u_0(x,t) = u(x,0) = (1+x), \quad (50)$$
$$u_1(x,t) = -\frac{\hbar(x+1)t^\alpha}{\Gamma(\alpha+1)}, \quad (51)$$
$$u_2(x,t) = \frac{\hbar(x+1)(\hbar t^\alpha \Gamma(\alpha+1) - \hbar\Gamma(2\alpha+1) - \Gamma(2\alpha+1))t^\alpha}{\Gamma(\alpha+1)\Gamma(2\alpha+1)}, \quad (52)$$

$$u_3(x,t) = -\frac{\hbar(x+1)t^\alpha}{\Gamma(\alpha+1)\Gamma(2\alpha+1)\Gamma(3\alpha+1)}\Big(\hbar^2 t^{2\alpha}\Gamma(\alpha+1)\Gamma(2\alpha+1) - 2\hbar^2 t^\alpha \Gamma(\alpha+1)\Gamma(3\alpha+1)$$
$$+\hbar^2 \Gamma(2\alpha+1)\Gamma(3\alpha+1) - 2\hbar t^\alpha \Gamma(\alpha+1)\Gamma(3\alpha+1) + 2\hbar\Gamma(2\alpha+1)\Gamma(3\alpha+1) \quad (53)$$
$$+\Gamma(2\alpha+1)\Gamma(3\alpha+1)\Big),$$

……………….

Now, we obtain the series solution from Eq. (14), using the Eq. (50-53). We obtain solution in closed form taking $\hbar = -1,$ and $\alpha = 1,$ we have

$$u(x,t) = (x+1)e^t. \qquad (54)$$

Which is exact solution of the problem when $\alpha = 1$.

**Example 4.4** In this example, we consider Eq. (1-2) with $n = 2$, $x_1 = x, x_2 = y$

$A_1(x,y,t) = x,$

$A_2(x,y,t) = 5y,$

$B_{1,1}(x,y,t) = x^2,$

$B_{1,2}(x,y,t) = 1,$

$B_{2,1}(x,y,t) = 1$

$B_{2,2}(x,y,t) = y^2$

$f(x) = x, \quad x, y \in \mathbb{R}.$

Applying the HATM, we have

$$R_m(\vec{u}_{m-1}) = \pounds[u_{m-1}] - \frac{(1-\chi_m)}{s}u(x,y,0) + \frac{2}{s^\alpha}\pounds[u_{m-1}] + \frac{y}{s^\alpha}\pounds[(u_{m-1})_y] - \frac{y^2}{s^\alpha}\pounds[(u_{m-1})_{yy}]$$
$$-\frac{3x}{s^\alpha}\pounds[(u_{m-1})_x] - \frac{2}{s^\alpha}\pounds[(u_{m-1})_{xy}] - \frac{x^2}{s^\alpha}\pounds[(u_{m-1})_{xx}], \qquad (55)$$

The mth-order deformation equation, $H(x,t) = 1$, we have

$$u_m(x,t) = \chi_m u_{m-1}(x,t) + \hbar\pounds^{-1}[R_m(\vec{u}_{m-1})]. \quad (m \geq 1), \qquad (56)$$

Using the Eq. (55-56), we have the following approximations

$$u_0(x,y,t) = u(x,y,0) = x, \qquad (57)$$

$$u_1(x,y,t) = -\frac{\hbar x t^\alpha}{\Gamma(\alpha+1)}, \qquad (58)$$

$$u_2(x,y,t) = \frac{\hbar x(\hbar t^\alpha \Gamma(\alpha+1) - \hbar\Gamma(2\alpha+1) - \Gamma(2\alpha+1))t^\alpha}{\Gamma(\alpha+1)\Gamma(2\alpha+1)}, \qquad (59)$$

$$u_3(x,y,t) = -\frac{\hbar x t^\alpha}{\Gamma(\alpha+1)\Gamma(2\alpha+1)\Gamma(3\alpha+1)}\Big(\hbar^2 t^{2\alpha}\Gamma(\alpha+1)\Gamma(2\alpha+1) - 2\hbar^2 t^\alpha \Gamma(\alpha+1)\Gamma(3\alpha+1)$$
$$+\hbar^2\Gamma(2\alpha+1)\Gamma(3\alpha+1) - 2\hbar t^\alpha \Gamma(\alpha+1)\Gamma(3\alpha+1) + 2\hbar\Gamma(2\alpha+1)\Gamma(3\alpha+1) \qquad (60)$$
$$+\Gamma(2\alpha+1)\Gamma(3\alpha+1)\Big),$$

…………..

Now, we obtain the series solution as from Eq. (14), using the Eq. (57-60). The solution in closed form taking $\hbar = -1,$ and $\alpha = 1,$ we have

$$u(x,t) = xe^t \qquad (61)$$

Which is exact solution of the problem when $\alpha = 1$.

**Example 4.5** Finally, we apply the HPTM on the nonlinear Fokker-Planck equation, consider Eq. (4) $n = 1, x_1 = x,$

$$A_1(x,t,u) = 4\frac{u}{x} - \frac{x}{3},$$

$$B_{1,1}(x,t,u) = u,$$

$$f(x) = x^2, \quad x \in \mathbb{R}.$$

Using the HATM, we have

$$R_m(\vec{u}_{m-1}) = \pounds[u_{m-1}] - \frac{(1-\chi_m)}{s}u(x,0) + \frac{8}{s^\alpha x}\pounds[\sum_{k=0}^{m-1}u_{m-1-k}(u_k)_x] - \frac{2}{s^\alpha}\pounds[\sum_{k=0}^{m-1}u_{m-1-k}(u_k)_{xx}] \qquad (62)$$

$$-\frac{2}{s^\alpha}\pounds[\sum_{k=0}^{m-1}(u_{m-1-k})_x(u_k)_x] - \frac{x}{3s^\alpha}\pounds[(u_{m-1})_x] - \frac{4}{x^2 s^\alpha}\pounds[\sum_{k=0}^{m-1}u_{m-1-k}u_k] - \frac{1}{3s^\alpha}\pounds[u_{m-1}],$$

The mth-order deformation equation, taking $H(x,t)=1$, we have

$$u_m(x,t) = \chi_m u_{m-1}(x,t) + \hbar \pounds^{-1}[R_m(\vec{u}_{m-1})]. \quad (m \geq 1), \qquad (63)$$

Using the Eq. (62-63), we have the following approximations

$$u_0(x,t) = u(x,0) = x^2, \qquad (64)$$

$$u_1(x,t) = -\frac{\hbar x^2 t^\alpha}{\Gamma(\alpha+1)}, \qquad (65)$$

$$u_2(x,t) = \frac{\hbar x^2 (\hbar t^\alpha \Gamma(\alpha+1) - \hbar \Gamma(2\alpha+1) - \Gamma(2\alpha+1))t^\alpha}{\Gamma(\alpha+1)\Gamma(2\alpha+1)}, \qquad (66)$$

$$u_3(x,t) = -\frac{\hbar x^2 t^\alpha}{\Gamma(\alpha+1)\Gamma(2\alpha+1)\Gamma(3\alpha+1)}\Big(\hbar^2 t^{2\alpha}\Gamma(\alpha+1)\Gamma(2\alpha+1) - 2\hbar^2 t^\alpha \Gamma(\alpha+1)\Gamma(3\alpha+1)$$
$$+ \hbar^2 \Gamma(2\alpha+1)\Gamma(3\alpha+1) - 2\hbar t^\alpha \Gamma(\alpha+1)\Gamma(3\alpha+1) + 2\hbar\Gamma(2\alpha+1)\Gamma(3\alpha+1) \qquad (67)$$
$$+ \Gamma(2\alpha+1)\Gamma(3\alpha+1)\Big),$$

…………..

Now, we obtain the series solution as from Eq. (14), using the Eq. (64-67). Now the solution in closed form taking $\hbar = -1,$ and $\alpha = 1$, we have

$$u(x,t) = x^2 e^t. \qquad (68)$$

## 5. Conclusion

The homotopy analysis transform method was successfully and effectively applied to solve the fractional linear and nonlinear Fokker-Plank equation. In this article we find exact solution of the initial value problem. The present work is easily shown that the series solution obtained by HATM converge rapidly. The solution obtained by HATM is in good agreement with $\alpha = 1$, and had a better explanation in fractional cases. The present study on fractional Fokker-Plank equation affirmed that homotopy analysis transform method is easy to apply and have less computational work with high accuracy. Therefore, this method may be applied to many complicated fractional nonlinear differential equations without any assumptions, transformations discretization and perturbation.